\documentstyle[aip,12pt,psfig,epsf]{article}

\def\la{\mathrel{\hbox{\rlap{\hbox{\lower4pt\hbox{$\sim$}}}\hbox{$<$}}}}
\def\ga{\mathrel{\hbox{\rlap{\hbox{\lower4pt\hbox{$\sim$}}}\hbox{$>$}}}}

\newcommand{\be}{\begin{eqnarray}}
\newcommand{\ee}{\end{eqnarray}}

\newcommand{\msol}{\ifmmode{{\rm M}_\odot}\else{M$_\odot$}\fi}
\newcommand{\foe}{\ifmmode{10^{51}}\else{$10^{51}$}\fi}

\newcommand{\xni}{\ifmmode{{\rm X}_{\rm Ni}}\else{X$_{\rm Ni}$}\fi}
\def\ang{\hbox{\AA}}
\def\Teff{\ifmmode{T_{\rm eff}}\else{\hbox{$T_{\rm eff}$} }\fi}
\def\Rzero{\ifmmode{R_0}\else{\hbox{$R_0$} }\fi}

  at 12.0 true pt 

\def\etal{{et al}}

\def\b{\beta}

\def\rout{\ifmmode{r_{\rm out}}\else\hbox{$r_{\rm out}$}\fi}
\def\tmax{\ifmmode{\tau_{\rm max}}\else\hbox{$\tau_{\rm max}$}\fi}
\def\tstd{\ifmmode{\tau_{\rm std}}\else\hbox{$\tau_{\rm std}$}\fi}
\def\vmax{\ifmmode{v_{\rm max}}\else\hbox{$v_{\rm max}$}\fi}
\def\muE{\ifmmode{\mu_{\rm E}}\else\hbox{$\mu_{\rm E}$}\fi} 
\def\pE{\ifmmode{p_{\rm E}}\else\hbox{$p_{\rm E}$}\fi} 
\def\bmax{\ifmmode{\b_{\rm max}}\else\hbox{$\b_{\rm max}$}\fi}
\def\kms{\hbox{$\,$km$\,$s$^{-1}$}}

\def\ang{\hbox{\AA}}

\def\Teff{\hbox{$\,T_{\rm eff}$} }

\def\rout{\hbox{$r_{\rm out}$} }

\def\chistd{\ifmmode{\chi_{\rm std}}\else\hbox{$\chi_{\rm std}$}\fi}

\def\doublet#1{\hbox{$ ^2$#1}}

\def\lstar{\ifmmode{\Lambda^*}\else\hbox{$\Lambda^*$}\fi} 
\def\Rop{\ifmmode{[R_{ij}]}\else\hbox{$[R_{ij}]$}\fi}

\def\Rji{\ifmmode{[R_{ji}]}\else\hbox{$[R_{ji}]$}\fi}
\def\Rstar{\ifmmode{[R_{ij}^*]}\else\hbox{$[R_{ij}^*]$}\fi}

\def\Rjistar{\ifmmode{[R_{ji}^*]}\else\hbox{$[R_{ji}^*]$}\fi}
\def\DRji{\ifmmode{[\Delta R_{ji}]}\else\hbox{$[\Delta R_{ji}]$}\fi}
\def\DRij{\ifmmode{[\Delta R_{ij}]}\else\hbox{$[\Delta R_{ij}]$}\fi}

\def\ns{\ifmmode{N_{\rm s}}          
        \else\hbox{$N_{\rm s}$}\fi}

\def\mat#1{{\bf #1}}     
\def\vek#1{{#1}}         

\newcount\eqcount
\def
  \nummer{
    \global\advance\eqcount by 1
    (\the\eqcount)
  }

\def
  \numadv{
    \global\advance\eqcount by 1
  }

\def
   \numout#1{
     (\the\eqcount #1)
  }

\def\ivek#1#2{\ifmmode{\vek{I}^{#1}_{#2}}
        \else\hbox{$\vek{I}^{#1}_{#2}$}\fi}

\def\tmat#1#2{\ifmmode{\mat{t}^{#1}_{#2}}
        \else\hbox{$\mat{t}^{#1}_{#2}$}\fi}
\def\rmat#1#2{\ifmmode{\mat{r}^{#1}_{#2}}
        \else\hbox{$\mat{r}^{#1}_{#2}$}\fi}
\def\bvek#1#2{\ifmmode{\beta^{#1}_{#2}}
        \else\hbox{$\beta^{#1}_{#2}$}\fi}

\def\lp{\ifmmode{\lambda^+_\tau}           
        \else\hbox{$\lambda^+_\tau$}\fi}
\def\lm{\ifmmode\lambda^-_\tau             
        \else\hbox{$\lambda^-_\tau$}\fi}

\def\etal{et al.}
\def\fep{\ifmmode{{\rm Fe II}}\else\hbox{Fe~II }\fi}

\begin{document}
\bibliographystyle{aip}

\title{Non-LTE Treatment of \fep in Astrophysical Plasmas}

\author{P.~H.~Hauschildt\\Dept. of
Physics and Astronomy\\Arizona State University\\Tempe, AZ 85287-1504\\
Email: {\tt yeti@sara.la.asu.edu}\\                                            
and\\
E. Baron\\Dept. of Physics and Astronomy\\
University of Oklahoma\\Norman, OK 73019-0225\\
Email: {\tt baron@monk.nhn.uoknor.edu}}
\date{\mbox{}}

\maketitle

\begin{abstract}
We describe our implementation of an extremely detailed model atom of
singly ionized iron for NLTE computations in static and moving
astrophysical plasmas. Our model atom includes 617 levels, 13675
primary permitted transitions and up to 1.2 million secondary
transitions. Our approach guarantees that the total iron opacity is
included at the correct wavelength with reasonable memory and CPU
requirements. We find that the lines saturate the wavelength space, such that
special wavelength points inserted along the 
detailed profile functions may be replaced with a statistical sampling
method. We describe the results of various test calculations for novae
and supernovae.
\end{abstract}

\section{Introduction}

Iron is the most common heavy element in the solar system. With 26
electrons it is quite difficult to treat theoretically as a quantum
mechanical system.  Because iron is so abundant and because of its low
ionization threshold of $\chi_I = 7.87$~eV, the lines of singly
ionized iron \fep are ubiquitous in astrophysical plasmas, from
stellar atmospheres to novae, supernovae, and quasars. Since it is so
complex, there are many terms and many multiplets in each
term. Previous efforts to perform NLTE calculations with \fep have had
to make rather crude approximations, e.g., lumping entire multiplets
together in a single ``super-level'' thereby reducing the model atom
to a manageable 30 to 50 levels. The energy spread within a multiplet
can correspond to a wavelength spread as large as 200\AA, so unless
something is done to correct for this, the opacity will not appear at
the correct wavelength. For modeling static plasmas such errors are
clearly unacceptable and further approximations, e.g., the usage of
opacity distribution functions (ODF's) in the radiative transfer
equation must be made. In rapidly expanding plasmas, e.g., those found
in novae and supernovae, the errors would be uncomfortably large and ODF's
cannot be applied. 
The method we describe in this paper does {\em not} use
any of the aforementioned approximations but includes the lines
directly in the radiative transfer equations at the correct
wavelength. Therefore, it can be applied to both moving and static
media, without either loss of accuracy or further assumptions.

 Another problem frequently encountered in classical multi-level NLTE methods 
is a badly conditioned rate matrix. This will significantly limit the number of 
individual levels that can be included in the calculation. The method that we 
introduce in this paper solves these potential problems by solving 
the statistical equations in a two step procedure, first solving well 
conditioned linear systems for the departure coefficients and afterwards 
solving the ionization and dissociation equations using an iterative 
procedure.  

In our effort to produce a multi-purpose multi-level NLTE stellar atmosphere 
code that is flexible enough to model a wide variety of astrophysical 
situations, we have developed an \fep model atom that includes all levels of 
a multiplet and, therefore, puts the opacity at the correct wavelength. 

\begin{figure}[t]
\caption[]{Grotrian Diagram of the \fep model atom that we use for the 
calculations presented here. All 617 level and 13675 b-b transitions which we 
include in NLTE are shown.}
\end{figure}

\section{Model Atom}

\subsection{Kurucz Data}
We have taken our model atom from the long term projectof R.~L.~Kurucz\cite{cdrom22}, 
to provide accurate atomic data for modeling stellar 
atmospheres is an invaluable service to the scientific community. For our 
current model atom we have kept terms to \doublet{H}, which corresponds to the 
first 29 terms of \fep. Within these terms we treat all observed levels that 
have observed b-b ``primary'' transitions with $\log{gf} > -3.0$, where g is 
the statistical weight of the lower level and f is the oscillator strength of 
the transition. This leads to a model atom with 617 levels and 13675 
``primary'' 
transitions which we treat in detailed NLTE;  we solve the full rate 
equations for all these levels including all radiative rates for the primary 
lines. In addition we treat the opacity and emissivity for the remaining 
nearly 1.2 million ``secondary'' transitions in NLTE.
 
\subsection{Photo-ionization Rates}
Detailed photo-ionization rates for \fep have yet to be published,
although this is one of the goals of the iron project. Thus, we have
taken the results of the Hartree Slater central field calculations of
Reilman \& Manson \cite{reilman79} to scale the ground state
photo-ionization rate and have then used a hydrogenic approximation
for the energy variation of the cross section. Although these rates
are only very rough approximations, they are useful for initial
calculations. In the conditions of the test cases we will consider in
this paper, the exact values of the b-f cross sections are not
important for the opacities themselves (which are dominated by known
b-b transitions of \fep and other species), but they have an influence
on the actual b-f rates.  This is, of course, unimportant for the
computational method which we use and the b-f cross sections can be
changed once better data become available.

\subsection{Collisional Rates}

For collisional rates we have had to make rather rough approximations
and, therefore, have approximated
bound-free collisional rates using the semi-empirical formula of
Drawin \cite{drawin61}. The bound-bound collisional rates are
approximated by the semi-empirical formula of Allen 
\cite{allen_aq}, while for  permitted transitions we use
Von~Regemorter's formula \cite{vr62}. While the collisional rates are important 
in stellar atmospheres with high electron densities, they are nearly 
negligible when compared to the radiative rates in the low density envelopes 
of novae and supernovae which we discuss here as test cases. Small collisional 
rates constitute a computationally much harder problem so that the tests we 
present here are good numerical tests of our computational method.

\section{Calculational Method}

The large number of transitions of the \fep ion that have to be included in 
realistic models of the \fep NLTE line formation require an efficient method for 
the numerical solution of the multi-level NLTE radiative transfer problem. As 
discussed below, the \fep model atom that we describe  includes, 
 more than 13000 individual lines. Classical approaches, e.g., the 
complete linearization or the Equivalent Two Level Atom methods, would be 
computationally prohibitive. In addition, we are also interested in modeling 
moving media, therefore, approaches like Anderson's multi-group
scheme\cite{and87} or 
extensions of the opacity distribution function method\cite{HubLan95} cannot be 
applied. Simple approximations like the Sobolev method are very inaccurate in 
problems in which lines overlap strongly and which have a significant continuum 
contribution (important for weak lines), as is the case for nova and SN 
atmospheres\cite{novaphys}. 

 We, therefore, use the multi-level operator splitting method described by 
Hauschildt\cite{casspap} (hereafter Paper II). This method solves the 
non-grey, spherically symmetric, special relativistic equation of radiative 
transfer in the co-moving (Lagrangian) frame using the operator splitting 
method described in Hauschildt\cite{s3pap} (hereafter Paper I). It has the 
advantages that (a) it is numerically highly efficient and accurate, (b) it 
treats the effects of overlapping lines and continua (including background 
opacities by lines and continua not treated explicitly in NLTE) 
self-consistently, (c) gives very stable and smooth convergence even in extreme 
cases (novae), and (d) it is not restricted to a certain application but can 
be applied to a wide variety of astrophysical problems. Details of the method 
are described in Paper I and II, so we give here only a discussion of the 
technical improvements necessary to make the NLTE treatment of very large 
model atoms more efficient.

\subsection{Full NLTE Treatment: Primary (strong) Lines}

Even with highly effective numerical techniques, the treatment of
possibly more than one million NLTE lines poses a significant
computational problem, in particular in terms of memory usage. In
addition, most of these lines are very weak and do not contribute
significantly to the radiative rates between their levels. However,
they may very well influence the radiation field in overlapping
stronger transitions and must therefore be included as background
opacity. Therefore, we separate the ``primary'' lines from the
``secondary'' lines by defining a threshold in $\log(gf)$, which can
be arbitrarily changed. Lines with gf-values larger than the threshold
are treated in detail, i.e., they are fully included as transitions in
the rate equations. In addition, we include special wavelength points
within the profile of the strong lines, whereas the weak lines are
treated by opacity sampling (see below). The weaker secondary
transitions are included as background NLTE opacity sources but are
not explicitly included in the rate equations. Their
cumulative indirect effect on the rates is included, as they are
considered in the solution of the radiative transfer equation. Note
that the distinction between primary and secondary transitions is just
a matter of convenience and technical feasibility. It is {\em not} a
restriction of our method or the computer code that we have developed
but is easily changed by altering the appropriate input files.  
As more powerful computers become available, all transitions can be 
handled as primary lines by simply changing the input files
accordingly.

 As a typical value, we use a threshold of $\log(gf) = -3$ so that lines with 
gf-values larger than this threshold are considered as primary lines and all 
other lines are considered as secondary lines. Note that we do not pose 
additional thresholds like the energy of the lower level of a line or the 
statistical weight of the lower level. However, we do include in the selection 
process only observed lines between observed levels in order to include only 
lines with well known gf-values. All predicted lines of Kurucz are included as 
``weak'' lines, see below. 

 Using this procedure to select our model atom, we obtain 13675 primary NLTE 
lines between the 617 levels included in NLTE. For every line we use 5 to 9 
wavelength points within its profile and for all lines the radiative rates and 
the ``approximate rate operators'' (see Paper II) are computed and included in 
the iteration process. We  discuss the effect of different treatment of 
the primary lines, e.g., by opacity sampling, in the results section.

\subsection{Approximate NLTE Treatment: Secondary Lines and LTE Levels}

The vast majority of the 1.2 million \fep lines in the database of Kurucz are 
either very weak lines or are predicted lines (sometimes between predicted or 
auto-ionizing levels). Although these weak lines are an important source for 
the overall opacity and the shape of the resulting ``pseudo-continuum'' (see 
below), they are not very important for the rate equations. The transitions 
between the bound states are dominated by a relatively smaller
number of  primary transitions, 
which we include individually in the radiative transfer and rate equation 
solution. 

 However, the ``haze'' of weaker ``secondary'' lines must be included as 
background opacity (and hence indirectly in the rate equations). This is also 
true for the numerous lines of species considered in LTE. Neglecting the line-
blanketing would lead to wrong results for the NLTE departure coefficients, 
see, e.g., Hauschildt \& Ensman\cite{snlisa} for a description of this effect 
found in supernova model atmospheres. 

 Therefore, we include the opacity of the secondary lines, defined as all 
available \fep lines that are not treated individually, as background opacity. 
Depending on the levels between which a weak transition takes place, we distinguish 
between \begin{enumerate} 
\item lines for which the lower level is an NLTE 
level but the upper level is an LTE level, 
\item lines for which the upper 
level is an NLTE level but the lower level is an LTE level, and, 
\item lines 
for which both levels are LTE levels. 
\end{enumerate} 
Here, an 'NLTE level' is 
a level that is explicitly treated in the NLTE rate equations whereas a 'LTE 
level' is not considered explicitly in the rate equations. In the first two 
cases we set the departure coefficient for the LTE level equal to the 
departure coefficient of the ground state,
whereas in the last case we use the same approach as for the 
lines of LTE species\cite{sn93jpapbig,cygpap,cas93pap} except 
that we use the ground state departure
coefficient to include the effects of over or under-ionization.
This approximate treatment of the secondary lines does not significantly 
influence the emergent spectra, as the secondary lines are by definition only 
relatively weak lines. 

\section{Results}

\subsection{LTE Tests}


 Before applying our method to simplified test cases, we have
performed the usual tests of the computer code and the stability of
the model atom. These include tests of the rate and statistical
equations with LTE values for either radiative rates alone, for
collisional rates alone, and for both LTE radiative rates and
collisional rates. We found that deviations of the departure
coefficients from their LTE value, $b_i=1$, is in all cases less than
$10^{-6}$ for a wide range in electron temperatures. These tests were
done by simply replacing the results of the monochromatic radiative
transfer calculations by their LTE values, but retaining the
approximate rate-operators in the calculations. This is very sensitive
to small numerical problems in the calculations, e.g., profile
normalization errors, which can lead to instabilities in model
calculations. Successful tests were very
important steps during the development of the large \fep model atom
and its implementation.

\subsection{The test models}

For the test calculations presented below, we have selected a number
of simplified nova and supernova model atmospheres. All models were
calculated using our general model atmosphere computer code {\tt
PHOENIX}, version 5.3, \cite{novaphys,casspap,s3pap,REpap,aliperf} which
implements the numerical method and \fep model atom as described
above. A more detailed description of the input physics and the code
itself can be found in the cited references.

 The nova model atmospheres are calculated using a power law density of the 
form $\rho(r)\propto r^{-N}$ with $N=3$. We have used a ballistic velocity law 
$v(r)\propto r$ with $\vmax=2000\kms$, typically found in nova atmospheres. 
Nova atmospheres are characterized by a large number of overlapping lines in 
the ultraviolet and, therefore, the Sobolev approximation {\em cannot} be used to 
model either the atmospheres or spectra of novae\cite{novaphys}. 
Novae are a very 
important test case because of the complicated structure of their atmospheres,
in particular due to their huge electron temperature gradients. We have 
selected two nova models as test cases: The first has an effective temperature 
of $\Teff=15000\,$K, which is representative of novae around their maximum 
light in the optical. In this model, the electron temperatures vary from 
$4500\,$K to more than $100000\,$K. The second nova test model has an 
effective temperature of $25000\,$K, it represents a later stage in the nova 
evolution. 

Our supernova test models correspond to the early phase of Type II
supernovae with $\Teff=7000\,$K and $9500\,$K and maximum expansion 
velocities of $8000\kms$ and $15000\kms$, respectively. 
As for the nova 
models, we have used a power law density, however, for the less extended 
shells of supernovae we have used a density exponent of $N=12$ in the
hot model and $N=9$ in the cool model. 
The hotter
model corresponds roughly to conditions much earlier than maximum
light, and the cooler model to just before maximum light. We will
discuss the NLTE effects of iron on post-maximum SN spectra and  Type I
supernovae atmospheres and spectra elsewhere. 
We have used solar compositions for both he nova and supernova models

\begin{figure}
\caption[]{\label{lte} Test of the \fep model atom. We compare 3 
different LTE nova test models in the optical (upper panel) and UV
spectra ranges. The full curve gives the spectrum calculated using a
pure LTE model, i.e., a model in which all lines were taken directly
from the LTE line list, including the \fep lines. The dashed curves
(which are practically identical to the full curve) are for a model in
which the NLTE lines were computed using the NLTE part of our computer
code with all departure coefficients set to unity.  This spectrum
includes both primary and secondary NLTE lines of \fep.  If we simply
omit the secondary \fep lines, we obtain the spectrum given by the
dotted curve.}
\end{figure}

 In order to check the implementation of the model atom and the effect of our 
distinction between primary and secondary NLTE lines, we have calculated some 
LTE test cases. Here, we display the results for the most sensitive test, a 
nova model atmosphere with $\Teff=15000\,$K. In Fig.~\ref{lte} we compare the 
spectra computed using a LTE model (all lines were taken from the LTE line 
list and the NLTE part of the code was switched off) with the spectra obtained 
replacing the LTE lines of our NLTE species with the lines computed in the 
NLTE part of the code but setting all departure coefficients to unity. This 
case is useful in order to understand the effects of the weak \fep lines on 
the radiation field and to test the correctness of the code. The two spectra 
are practically identical, the relative differences are less than 0.1\%. 
Even though in these tests the iron abundance is low,  it
contributes significantly to the opacity and large differences would
significantly effect the output spectra. 

In addition, we have calculated a spectrum in which we have artificially 
neglected the secondary \fep lines and included only the 13675 primary lines 
of \fep (all NLTE lines of the other species as well as the lines of the 
species treated on LTE were 
retained) and plotted it in the dotted curve. Even in this case the spectra 
are nearly identical, we find only minor differences at some UV wavelengths, 
while in the optical spectral range the differences are localized in a few 
relatively weak transitions. This demonstrates that our \fep model atom is 
large enough and includes enough lines as primary transitions to model the 
spectrum and the total opacity of \fep. The secondary lines are a relatively 
small correction and their influence on the spectrum and the lines formation 
is, as intended, of secondary importance.

\begin{figure} 
\caption[]{\label{conv} Convergence properties of the departure coefficients.
The relative corrections $|\Delta b_i|/b_i$ of the ground state departure 
coefficients are shown as functions of the iteration number. 
The corrections for a number of different
layers in the atmospheres are plotted and the symbols indicate the ``standard
optical depth'' $\tstd$ (see text).  
%
The results are for a nova model with $\Teff=15000\,$K and  a
diagonal $\Rstar$-operator has been used for all species. The results
for the OS/ALI iterations are indicated by the lines connecting the
symbols. For comparison we show also the results of the
$\Lambda$-iterations with unconnected symbols.}
\end{figure}

\subsection{Convergence Properties}

 To test the convergence properties of the multi-level radiative
transfer and rate equations, we have calculated a number of simplified
test models.  In the nova and supernova test calculations we find that
the convergence of the departure coefficients is to a good
approximation linear and avoids the stabilization typically found in
the $\Lambda$-iteration. We show in Fig.~\ref{conv} the relative
corrections to the ground state departure coefficients $b_1$ of \fep
as functions of the iteration number at a number of optical depth
points. We define ``standard optical depth scale'' $\tstd$ as the
optical depth in the b-f continuum at $\lambda=5000\ang$.  The departure
coefficients are defined as the ratio of the NLTE occupation numbers
and their LTE values, computed using the NLTE occupation number of the
ground state of the next higher ion (see Paper II for details). The
two outermost optical depths are outside the line forming region
whereas the inner $\tau$ points are inside the line forming
regions. The last optical depth point plotted is inside the
thermalization depth of \fep. In addition to the 
operator-splitting---accelerated lambda iteration (OS/ALI) results
(indicated by the
line connecting the symbols) we have also plotted the results of the
simple $\Lambda$ iteration.

The \fep converges well and does not show any
problematic behavior. The convergence is better at larger optical
depths because the exact diagonal $\Rstar$-operator (see Paper II) is a better
approximation to the discrete rate operator at large optical
depths. In smaller and intermediate optical depths the exact
tri-diagonal $\Rstar$ would lead to much better convergence, but for a
model atom as large as the \fep that we use here, it requires
significantly more memory and CPU time per iteration. Thus, in the
calculations reported in this paper, we have used only the exact
diagonal $\Rstar$-operator, which leads to a somewhat lower
convergence rate (although the exact tri-diagonal $\Rstar$-operator is
available in our computer code). At very small optical depths the
convergence rates are somewhat better because the radiation field is
dominated by the background field emergent from the line forming
region and the effects of local radiation fields are much smaller (in
effect, this is the transition into the nebular region surrounding a
typical nova atmosphere).

We have made the situation more complicated  by including a 
number of other species in the NLTE calculation, as indicated in 
Fig.~\ref{bis}. 
These species 
were not directly coupled to the \fep model atoms in order to avoid the 
problematic case of unknown couplings between different species. Furthermore, 
we have only included direct couplings between transitions in the \fep model 
atoms in order to save computer time (although the full set of couplings 
between lines can easily be included in the calculations at the cost of  
CPU time). As demonstrated in Papers I and II, much 
better convergence rates are attained if every possible coupling is included 
in the calculations and if convergence acceleration methods are used.
The cases we discuss here should be close to worst case 
scenarios. That the convergence rates are still acceptable shows the 
robustness of our numerical method.

\begin{figure}
\caption[]{\label{bis} Departure coefficients of the ground state of a number of NLTE species 
as functions of standard optical depth.
The different symbols indicate the different species as shown in the legend of 
the figure.}
\end{figure}

\begin{figure}
\caption{\label{bis.sn} 
Departure coefficients of the ground state of
\fep\ for the hot and cool supernova models
as functions of standard optical depth.}
\end{figure}

\subsection{Departures from LTE}

\subsubsection{Ionization Changes}
 In Fig.~\ref{bis} we show the departure coefficients $b_1$ of the
ground states of our NLTE species as functions of $\tstd$. In the nova
models, the run of $b_1$ with optical depth is relatively complicated
throughout the line and continuum forming regions. The ground state
departure coefficients of \fep are less than unity in most regions of
the atmosphere, indicating an over-ionization of \fep relative to the
LTE situation. At intermediate optical depths ($\tstd \sim 1 - 100$)
the \fep $b_1$ are however, larger than unity. This complicated
behavior is mirrored by the other NLTE species, in particular by
O~I. For most species, the change from over to under-ionization occurs
over a larger range in optical depth and the ``pseudo-nebular'' outer
region where $b_1<1$ is relatively smaller.

Fig.~\ref{bis.sn} displays the ground state departure coefficient for
our hot and cool supernova models. For the hot model the effect of
NLTE is over-ionization throughout the line forming region. The cool
model displays the complicated behavior seen in the nova models.

\subsubsection{NLTE effects of the \fep lines}

Since \fep is such a complicated atom with so much coupling between
levels, it would not be surprising if the main effect of NLTE was
over-ionization with the internal population density {\em ratios}
being close to LTE once the
over-ionization is taken into account. In Figure~\ref{biratio} we plot
the ratio of the departure coefficients 
to that of the ground
state for the levels $z\,^6D^o$, $b\,^4P$, and $z\,^4D^o$ which
correspond to the upper level of the $\lambda 2600$ UV1 transition,
and the
lower and upper levels for the $\lambda 4233$ optical transition,
respectively. 

If the main effect of NLTE was that the hot radiation
field over-ionizes  \fep, then we would expect that the departure
coefficient ratio should become constant independent of depth and that
the constant would be proportional to $\exp({-\chi/kT_R})$, where $T_R$
is the characteristic temperature of the radiation field, i.e. that a
two-fluid approach would be valid. For the cool
model, this is never the case and the populations of the levels are
far from any kind of thermal equilibrium value. This is not surprising given
the {\em highly\/} non-Planckian character of the radiation field.
Only for the very low optical depths of the hot model does over-ionization
appear to be the main effect. In the line formation region, however
the behavior is quite complex. 

\begin{figure}
\caption[]{\label{biratio}
The ratio of the departure coefficients to the ground
state departure coefficient for our cool (lower panel)  and hot (upper
panel)  test supernova models. The
lines correspond to the $z\,^6D^o$ level (the upper level for the
$\lambda 2600$ UV1 transition), $z\,^4D^o$ level (the upper level for
the $\lambda 4233$ optical transition), and the $b\,^4P$ level (the
lower level for the $\lambda 4233$ optical transition).}
\end{figure}

\subsection{Effects of \fep NLTE on the emitted spectrum}

\begin{figure}
\caption[]{\label{novanlte} The effect of \fep NLTE on nova spectra. The 
full curves give the synthetic spectrum of a model atmosphere calculated 
including the \fep NLTE effects whereas the dotted curve gives the 
spectrum calculated using LTE population number for \fep. All models 
include, in addition, NLTE effects of H~I, He~I, O~I, Na~I, Ne~I, Mg~II, and 
Ca~II, and line blanketing by other metals. The structures of the 
atmospheres are identical. The effective temperature of the model
atmosphere is $\Teff=15000\,$K.}
\end{figure}

\begin{figure}
\caption{\label{novanlteb} Same as Fig.~\protect\ref{novanlte}, but for $\Teff=25000\,$K}
\end{figure}

\subsubsection{Nova models}

 The effects of \fep NLTE on the spectra emitted by novae are very large. We 
demonstrate this in Figs.~\ref{novanlte} and \ref{novanlteb} by comparing the 
spectra for two nova model atmospheres ($\Teff=15000\,$K and $25000\,$K) 
computed assuming LTE population numbers for \fep and using the full NLTE 
treatment for the 617 level \fep model atom.  The most obvious effect on the 
optical spectra, for both effective temperatures, is that \fep NLTE reduces the 
emission strength of H$\beta$ (and the higher members of the Balmer series for 
$\Teff=25000\,$K). This is due to a reduced optical depth in the Lyman lines, 
caused by under-population and correspondingly weaker \fep lines in the 
wavelength range from $900$ to $1250\ang$. This alters the departure 
coefficients for hydrogen and thus changes the profiles of the Balmer lines. 
The effect of \fep NLTE on the optical \fep lines seems relatively small, in 
the 10\% range. However, this is important for iron abundance analyses in 
novae, as this is roughly equivalent to a factor of two change in the Fe 
abundance. 

 The situation is very different in the UV spectral range. For the
$\Teff=15000\,$K model the NLTE effects of \fep are not dramatic, in
particular the \fep--Mg~II blend at $2800\ang$ is practically
unchanged. The feature at $2640\ang$ (actually a gap between strong
\fep lines, see Hauschildt \etal\cite{novaphys}) is slightly
``weaker'' (i.e., the opacity is enhanced) in the \fep--NLTE model
when compared to the \fep--LTE model. For the nova atmosphere with
$\Teff=25000\,$K, however, the changes are significantly larger. The strong
\fep lines which are visible in the \fep--LTE model at $\lambda\approx
2450,$ $2650$ (at this effective temperature, this feature is a
cluster of \fep emission lines), and $2770\ang$ are practically wiped
out in the \fep--NLTE model. This is caused by a strong
under-population of their levels and a large over-ionization of \fep
in the NLTE case. The \fep--NLTE model is in much better agreement
with actually observed nova spectra (Schwarz \etal, in preparation)
than the previous LTE model spectra. In addition to this very
important effect, the \fep--NLTE model does not show some of the broad
absorption features found in the \fep--LTE spectrum around $1620\ang$.

\subsubsection{Supernova models}

Figures~\ref{snnlte} and \ref{snnlteb} show the effect of \fep NLTE on the spectra emitted by the 
supernova models. With the high velocities characteristic of  early supernova 
atmospheres, the individual features are highly blended. None of the 
effects on the line shapes is as dramatic as in the case of novae. It is 
interesting, however,
that for the cooler models the largest effects are in the optical 
with noticiable changes in H$\beta$ and the \fep features between 4000 and $5000\ang$. 
For the hotter model, the largest changes occur for the \fep lines in the UV. 
This is somewhat similar to the nova model. 
In Type Ia supernovae where iron is a major atmospheric constituent the 
effects are much larger and we will discuss them in detail in a specialized 
paper (Nugent \etal, in preparation). 

\begin{figure} 
\caption[]{\label{snnlte} The effect of \fep NLTE on supernova 
spectra. The dotted curves give the synthetic spectrum of the cool 
model atmosphere calculated including the \fep NLTE effects whereas the solid 
curve gives the spectrum calculated using LTE population number for \fep. All 
models include, in addition, NLTE effects of H~I, He~I, O~I, Na~I, Ne~I, 
Mg~II, and Ca~II, and line blanketing by other metals. The structures of the 
atmospheres are identical.} 
\end{figure} 

\begin{figure}
\caption{\label{snnlteb} Same as Fig.~\protect\ref{snnlte}, but for the hot supernova model.}
\end{figure}

 These results show that NLTE effects of \fep are (a) not only important for the 
formation of the \fep lines themselves but (b) in addition they can also have an 
important indirect effect on the formation of the hydrogen lines (at least). 

\begin{figure}
\caption{\label{sampling}Effect of sampling on nova spectra.}
\end{figure}

\subsection{Sampling vs special wavelength points}

 The treatment of primary \fep transitions by inserting 5 to 7
wavelength points per NLTE line is very accurate, but also very costly
in terms of CPU time (its effect on storage is negligible). In order
to reduce CPU requirements, we have also implemented an alternate
method of treating primary NLTE lines, opacity sampling. The majority
of the primary \fep transitions are in the spectral range from
$900\ang$ to $3500\ang$. In this range, the transitions of \fep, and
other species, lie very close to one another and they overlap
strongly. Hence, it is not necessary to insert 5 wavelength points
per primary line, if the basic wavelength grid is already fine enough
to resolve the line opacity and the radiative rates accurately
enough. In the case of moving envelopes of hot stellar objects, such
as novae and supernovae, this is actually made simpler by the presence
of the $\partial/\partial \lambda$ term in the Lagrangian frame
radiative transfer equation. We have tested this method for a smaller
\fep model atom (472 level, 4500 primary transitions) in order to save
computer time. The resulting synthetic spectra are practically
identical in the test models we have calculated, as shown in
Fig.~\ref{sampling}. We use the sampling procedure only for \fep lines
between $900$ and $3500\ang$, all other primary \fep NLTE lines, as
well as {\em all} lines of the other NLTE species, are computed using
5 to 7 specially inserted wavelength points per lines. This procedure
reduces the number of wavelength points by about a factor of 2 to 3
and reduces the totally CPU time for the model construction also by
factors of 2 to 3 (the time required for a model iteration is directly
proportional to the number of wavelength points). Another big
advantage of the NLTE opacity sampling method is that additional
species with many lines in the sampling region (e.g., cobalt, nickel,
or titanium), will {\em not} require additional wavelength points,
resulting in a very small increase in computer time as we include more
complex NLTE model atoms.

\section{Conclusions}

We have implemented a full NLTE treatment for an extremely detailed model atom 
of \fep. With 617 levels, 13675 primary and 1.2 million secondary transitions 
in our \fep model atom, these calculations are to our knowledge the most 
detailed NLTE \fep calculation to date. By making use of the properties of the 
OS/ALI method developed in Papers I and II, we are  able to run our 
calculations on small workstations (a complete model can be calculated 
in less than 2 days on an 
IBM RS/6000 320 workstation, in about 6 hours CPU time on 
a RS/6000 580, and in about 1.5 hours CPU time on a Cray C90). We have been able 
to minimize memory and CPU requirements by using a diagonal $\Rstar$-operator, 
but the exact tri-diagonal operator is available for use on larger and faster 
computers. 

Our radiative transfer method is fully relativistic to all orders in
$v/c$ and no simplifying assumptions such as the Sobolev approximation
have been used. In addition, the detail of our model atom guarantees
that the opacity occurs at the correct wavelength so that our
calculations are applicable to both static and moving media.

We have been able to further reduce CPU demands by replacing a significant 
fraction of specially inserted wavelength points for the primary transitions
with a statistical sampling technique. This
is possible due to the detail of our model atom which ensures that
wavelength space is saturated with primary lines. 
Hence, the inclusion of additional
interesting iron peak ions will not substantially increase
computational requirements.

We have presented the results of test calculations in novae and
supernovae and find that the level populations in the line forming
region are quite complex and cannot be simply approximated by a
two-fluid approach which assumes that the levels are in internal equilibrium
with a hot radiation field and not with the matter. Although this approach 
would account correctly for the changes in the \fep ionization equilibrium, it 
cannot correctly model the line formation of \fep\unskip.  

 The effects of \fep NLTE on the spectra of nova and supernova atmospheres are 
large, in particular for the hotter nova models. With the method we have 
describe in this paper, it is possible to treat the \fep NLTE line formation 
in great detail, and this is required in order to be able to correctly and 
meaningfully interpret nova and SN spectra. In forthcoming papers we will 
report the results of detailed analyses of nova and SN spectra using the 
approach we have described here. 

\medskip
\noindent{\em Acknowledgements:}
It is a pleasure to thank F. Allard, D. Branch, P. Nugent, S. Shore,
and S. Starrfield for stimulating discussions.
This work was supported 
in part by a NASA LTSA grant to Arizona State University, and by NASA grant 
NAGW-2999. Some of the calculations presented in this paper  were 
performed at the San Diego Supercomputer Center (SDSC), supported by the 
NSF, and at the NERSC, supported by the U.S. DoE, we thank them for a 
generous allocation of computer time.

\bibliography{yeti,supernovae,opacity,novae,mdwarf}

\end{document}